\pdfoutput=1
\RequirePackage{ifpdf}
\ifpdf % We~are running pdfTeX in pdf mode
\documentclass[pdftex]{sigma}
\else
\documentclass{sigma}
\fi

\numberwithin{equation}{section}

\begin{document}
\allowdisplaybreaks

\newcommand{\arXivNumber}{1902.08598}

\renewcommand{\PaperNumber}{124}

\FirstPageHeading

\ShortArticleName{Further Results on a Function Relevant for Conformal Blocks}

\ArticleName{Further Results on a Function Relevant \\ for Conformal Blocks}

\Author{Vincent COMEAU~$^\dag$, Jean-Fran\c{c}ois FORTIN~$^\ddag$ and Witold SKIBA~$^\S$}

\AuthorNameForHeading{V.~Comeau, J.-F.~Fortin and W.~Skiba}

\Address{$^\dag$~Department of Physics, McGill University, Montr\'eal, QC H3A 2T8, Canada}
\EmailD{\href{mailto:vincent.comeau@mail.mcgill.ca}{vincent.comeau@mail.mcgill.ca}}

\Address{$^\ddag$~D\'epartement de Physique, de G\'enie Physique et d'Optique,\\
\hphantom{$^\ddag$}~Universit\'e Laval, Qu\'ebec, QC G1V 0A6, Canada}
\EmailD{\href{jean-francois.fortin@phy.ulaval.ca}{jean-francois.fortin@phy.ulaval.ca}}

\Address{$^\S$~Department of Physics, Yale University, New Haven, CT 06520, USA}
\EmailD{\href{witold.skiba@yale.edu}{witold.skiba@yale.edu}}

\ArticleDates{Received July 07, 2020, in final form November 24, 2020; Published online November 30, 2020}

\Abstract{We present further mathematical results on a function appearing in the conformal blocks of four-point correlation functions with arbitrary primary operators. The $H$-function was introduced in a previous article and it has several interesting properties. We prove explicitly the recurrence relation as well as the $D_6$-invariance presented previously. We also demonstrate the proper action of the differential operator used to construct the $H$-function.}


\Classification{33C70; 33C65; 33C90; 81T40}

\section{Introduction}\label{SecIntro}

Conformal field theories (CFTs) play an important role in modern physics. The introduction of the full conformal algebra constrains non-trivially $N$-point correlation functions. For example, two-point correlation functions are completely determined by conformal invariance while three-point correlation functions are settled in terms of a finite number of coefficients. This can be seen as originating from the existence of a convergent operator product expansion (OPE)~\cite{Mack:1976pa}. Moreover, using the OPE twice in four-point correlation functions leads to conformal blocks which represent the contributions of exchanged primary operators to the four-point correlation functions. Using associativity in $N$-point correlation functions further constrains the OPE, leading to the crossing symmetry of the four-point correlation functions which can be used to restrict the unknown conformal dimensions and OPE coefficients \cite{Ferrara:1973yt,Polyakov:1974gs}. Recent work in spacetime dimensions larger than two resulted in some explicit expressions for certain specific conformal blocks \cite{Dolan:2000ut,Dolan:2003hv}, and the conformal bootstrap in spacetime dimensions larger than two has also been implemented numerically with impressive results \cite{Rattazzi:2008pe}, see \cite{Poland:2018epd} for a review of recent results on conformal bootstrap.

The computation of conformal blocks in spacetime dimensions larger than two is not straightforward. Several approaches to this challenging problem have been developed, for example shadow formalism \cite{Ferrara:1972uq,Ferrara:1972xe,SimmonsDuffin:2012uy}, recurrence relations \cite{Penedones:2015aga, Zamolodchikov:1985ie,Zamolodchikov:1987}, harmonic analysis \cite{Dobrev:1977qv}, weight shifting \cite{Karateev:2017jgd}, and integrable systems \cite{Isachenkov:2017qgn, Schomerus:2017eny,Schomerus:2016epl}. Although the conformal blocks are technically determined by conformal symmetry, they are better understood from the embedding space where the conformal generators act linearly \cite{Dirac:1936fq,Mack:1969rr,Weinberg:2010fx,Weinberg:2012mz}. For example, the OPE has been studied utilizing the embedding space formalism in \cite{Ferrara:1971zy,Ferrara:1973eg,Ferrara:1971vh,Ferrara:1972cq,Fortin:2016lmf,Fortin:2016dlj}. In~\cite{Fortin:2016lmf}, it was shown how to employ the OPE in the embedding space formalism to compute the scalar conformal block. In~\cite{Fortin:2016dlj}, the results of \cite{Fortin:2016lmf} were used to find a very general function, the $H$-function, that appears in general conformal blocks containing fields in more complicated representations. Using a Rodrigues equation for the $H$-function, it was shown that it satisfies a recurrence relation and a~specific symmetry property leading to invariance under the dihedral group of order~$12$.

In this paper, an explicit expression for the $H$-function is used to show directly that the recurrence relation and the symmetry property are indeed satisfied. After briefly reviewing the definition of the $H$-function and analyzing its convergence in Section~\ref{SecF}, several new expressions for this function are obtained in Section~\ref{SecExp}. Section~\ref{SecRec} gives a proof of the recurrence relation and the symmetry property using various expressions explored in the previous section. Finally, in Section~\ref{SecDiff} the action of the differential operator is found and it is shown how to use it to compute the $H$-function constructively.

\section{Functions}\label{SecF}

In this section we review the $G$- and $H$-functions as well as the associated differential operators introduced in~\cite{Fortin:2016dlj}. We also study the convergence of the $H$-function using Horn's technique.

\subsection{Power series}

It is a well-known fact that the conformal blocks for scalar exchange in four-point correlation functions of four scalar primary fields in arbitrary spacetime dimension $d$ are related to the function $G_d^{(q;r;t)}(u,v)$. This function can be expressed as a double sum over powers of the variables $x=u/v$ and $y=1-1/v$,
\begin{equation}
 G_d^{(q;r;t)}(u,v)=\sum_{m,n\geq0}\frac{(-q,-t)_m}{(r-t+1-d/2)_mm!}\frac{(r,r-t+q)_{m+n}}{(r-t)_{2m+n}n!}x^my^n, \label{EqG}
\end{equation}
where $u$ and $v$ are the conformal cross-ratios and $(a)_n$ represents the rising Pochhammer symbol with $(a,b,\dots)_n=(a)_n(b)_n\cdots$. Moreover, \eqref{EqG} can be expressed in terms of the hypergeometric function $G(\alpha,\beta,\gamma,\delta;x,y)$ of Exton \cite{Dolan:2003hv,0305-4470-28-3-017,Fortin:2016dlj} as
\begin{displaymath}
 G_d^{(q;r;t)}(u,v)=G(r,r-t+q,r-t+1-d/2,r-t;x,y),
\end{displaymath}
where $q$, $r$ and $t$ are related to the conformal dimensions of the five scalar primary fields appearing in the conformal blocks.

In \cite{Fortin:2016dlj}, it was argued that more general conformal blocks are given by linear combinations of the following function,
\begin{gather}
H_d^{(p,q;r;s,t)}(u,v) =\sum_{m,n\geq0}P_d^{(p,q;r;s,t)}(m,n)\nonumber\\
\hphantom{H_d^{(p,q;r;s,t)}(u,v) =}{}\times
\frac{(r,r-s+p,r-t+q)_{m+n}}{(r-s+1-d/2,r-t+1-d/2)_m(r-s,r-t)_{2m+n}n!}x^my^n,
\nonumber\\
P_d^{(p,q;r;s,t)}(m,n) =\sum_{i,j\geq0}\frac{(-i)_j}{i!j!(m-i)!}(-p,r-t+1-d/2+m-i)_i\nonumber\\
\hphantom{P_d^{(p,q;r;s,t)}(m,n) =}{}\times (-s-q+m-i+j,r-t+2m+n-i+j)_{i-j}\label{EqH}\\
\hphantom{P_d^{(p,q;r;s,t)}(m,n) =}{}\times (-q)_{m-i+j} (-t)_{m-i+j} (r-s+m+n+i,r-s+p+1-d/2)_{m-i}.\nonumber
\end{gather}
The functional form \eqref{EqH}, as well as several contiguous relations and the symmetry properties $H_d^{(p,q;r;s,t)}(u,v)=H_d^{(q,p;r;t,s)}(u,v)$ and $H_d^{(p,q;r;s,t)}(u,v)=H_d^{(p,t;r-t+q;s-t+q,q)}(u,v)$, which generate the dihedral group $D_6$ of order $12$, were obtained by using the definitions~\eqref{EqG} and~\eqref{EqH} in terms of differential operators,
\begin{gather}
G_d^{(q;r;t)}(u,v) =\frac{\left(\frac{u}{v}\right)^{-(r-t+q)}\mathcal{D}_{(u,v)}^q\left(\frac{u}{v}\right)^{r-t}v^r}{(-2)^q(r-t,r-t+1-d/2)_q},\nonumber\\
H_d^{(p,q;r;s,t)}(u,v) =\frac{\left(\frac{u}{v}\right)^{-(r-s+p)}\mathcal{D}_{(u,v)}^p\left(\frac{u}{v}\right)^{t-s-q}\mathcal{D}_{(u,v)}^q \left(\frac{u}{v}\right)^{r-t}v^r}{(-2)^{p+q}(r-s,r-s+1-d/2)_p(r-t,r-t+1-d/2)_q}\nonumber\\
\hphantom{H_d^{(p,q;r;s,t)}(u,v)}{} =\frac{\left(\frac{u}{v}\right)^{-(r-s+p)}\mathcal{D}_{(u,v)}^p\left(\frac{u}{v}\right)^{r-s}G_d^{(q;r;t)}(u,v)}{(-2)^p(r-s,r-s+1-d/2)_p}.\label{EqGH}
\end{gather}
In the equation above, the second-order differential operator $\mathcal{D}_{(u,v)}$, as well as two related first-order differential operators $\mathcal{D}_{(u)}$ and $\mathcal{D}_{(v)}$, are defined as
\begin{gather*}
\mathcal{D}_{(u,v)}=(-2)\big\{u^3\partial_u^2+u^2(u+v-1)\partial_u\partial_v+u^2v\partial_v^2 \nonumber\\
\hphantom{\mathcal{D}_{(u,v)}=}{} -\big(\tfrac{d}{2}-2\big)u^2\partial_u+u\big[u+\big(\tfrac{d}{2}-1\big)(1-v)\big]\partial_{v}\big\},\nonumber\\
\mathcal{D}_{(u)}=-2u\partial_u-(u+v-1)\partial_v,\qquad \mathcal{D}_{(v)}=u(u-v-1)\partial_u+v(u-v+1)\partial_v,%\label{EqD}
\end{gather*}
and satisfy the algebra
\begin{gather*}
 \big[\mathcal{D}_{(u)},\mathcal{D}_{(v)}\big]=\mathcal{D}_{(u)}-\mathcal{D}_{(v)}, \qquad\big[\mathcal{D}_{(u)},\mathcal{D}_{(u,v)}^h\big]=-2h\mathcal{D}_{(u,v)}^h, \qquad\big[\mathcal{D}_{(v)},\mathcal{D}_{(u,v)}^h\big]=-2h\mathcal{D}_{(u,v)}^h.
 \end{gather*}

\subsection{Convergence}

Before we proceed further we investigate convergence of the $H$-function introduced in \eqref{EqH}. In its present form, the $H$-function is not a standard hypergeometric function of two variables as the coefficients of its power series contain $P_d^{(p,q;r;s,t)}(m,n)$, a sum of hypergeometric-like terms. We note that both sums in the definition of $P_d^{(p,q;r;s,t)}(m,n)$ terminate for generic values of parameters $p$, $q$, $r$, $s$ and $t$.

Convergence of multi-variable hypergeometric functions can be deduced using the Horn's method \cite{Horn1889}. Therefore, we represent the $H$-function in terms of a generalized four-variable hypergeometric function
\begin{gather}
L\left[\left. \begin{matrix}\alpha_1,\alpha_2,\alpha_3;\beta;\gamma;\delta_1,\delta_2;\epsilon\\a,b,c,d,e\end{matrix}\right|x_1,x_2,x_3,x_4\right]
\nonumber\\
\qquad{}=\sum_{n_1,n_2,n_3,n_4\geq0}\frac{(\alpha_1,\alpha_2,\alpha_3)_{n_1+n_2+n_3+n_4}(\beta)_{n_1+n_2+n_3} (\gamma)_{n_1+n_2}(\delta_1,\delta_2)_{n_2+n_3}(\epsilon)_{n_3}}{(a)_{2n_1+2n_2+n_3+n_4} (b)_{n_1+2n_2+2n_3+n_4}(c)_{n_1+n_2+n_3}(d)_{n_2+n_3}(e)_{n_3}}\nonumber\\
\qquad\quad{}\times \frac{x_1^{n_1}x_2^{n_2}x_3^{n_3}x_4^{n_4}}{n_1!n_2!n_3!n_4!},
\label{EqL}
\end{gather}
such that
\begin{gather*}
 H_d^{(p,q;r;s,t)}(u,v)\\
 \qquad{} =L\left[\left. \begin{matrix}r,r-s+p,r-t+q;-s-q;-p;-q,-t;r-s+p+1-d/2\\r-s;r-t;r-s+1-d/2;-s-q;r-t+1-d/2\end{matrix} \right|x,-x,x,y\right].
\end{gather*}

\subsubsection{Horn's technique}

In order to generalize the Horn's technique (see \cite{Horn1889,Srivastava1972,Srivastava1985}) to four-variable functions
\begin{displaymath}
 F=\sum_{n_1,n_2,n_3,n_4\geq0}A_{n_1,n_2,n_3,n_4}x_1^{n_1}x_2^{n_2}x_3^{n_3}x_4^{n_4},
\end{displaymath}
we define
\begin{displaymath}
 f_i(n_1,n_2,n_3,n_4)=\frac{A_{n_1+\delta_{1i},n_2+\delta_{2i},n_3+\delta_{3i},n_4+\delta_{4i}}}{A_{n_1,n_2,n_3,n_4}},
\end{displaymath}
and
\begin{gather*}
u_i(\mu_1,\mu_2,\mu_3,\mu_4)=\big|\lim_{\lambda\to\infty}f_i(\mu_1\lambda,\mu_2\lambda,\mu_3\lambda,\mu_4\lambda)\big|^{-1},\\
R_i=u_i(\delta_{1i},\delta_{2i},\delta_{3i},\delta_{4i}).
\end{gather*}
The region of convergence $|x_i|<r_i$ corresponds to the intersection of the following sets
\begin{gather*}
S_1=\big\{(r_1,r_2,r_3,r_4)\,|\,\land_i[0<r_i<u_i(\delta_{1i},\delta_{2i},\delta_{3i},\delta_{4i})=R_i]\big\},\\
S_{ij}=\big\{(r_1,r_2,r_3,r_4)\,|\,\forall\,(\mu_i,\mu_j)\in\mathbb{R}_+^2\colon \lor_{k=i,j}[0<r_k<u_k]\big\},\\
S_{ijk}=\big\{(r_1,r_2,r_3,r_4)\,|\,\forall\,(\mu_i,\mu_j,\mu_k)\in\mathbb{R}_+^3\colon \lor_{l=i,j,k}[0<r_l<u_l]\big\},\\
S_{1234}=\big\{(r_1,r_2,r_3,r_4)\,|\,\forall\,(\mu_1,\mu_2,\mu_3,\mu_4)\in\mathbb{R}_+^4\colon \lor_i[0<r_i<u_i]\big\},
\end{gather*}
i.e.,
\begin{gather*}
 D=S_1\cap S_{12}\cap S_{13}\cap S_{14}\cap S_{23}\cap S_{24}\cap S_{34}\cap S_{123}\cap S_{124}\cap S_{134}\cap S_{234}\cap S_{1234},
\end{gather*}
where $D$ is the convergence region.

\subsubsection{Region of convergence}

Applying the results from the proceeding section to \eqref{EqL}, we find
\begin{gather}
u_1=\frac{\mu_1(2\mu_1+2\mu_2+\mu_3+\mu_4)^2(\mu_1+2\mu_2+2\mu_3+\mu_4)}{(\mu_1+\mu_2)(\mu_1+\mu_2+\mu_3+\mu_4)^3},\nonumber\\
u_2=\frac{\mu_2(2\mu_1+2\mu_2+\mu_3+\mu_4)^2(\mu_1+2\mu_2+2\mu_3+\mu_4)^2}{(\mu_1+\mu_2)(\mu_2+\mu_3)(\mu_1+\mu_2+\mu_3+\mu_4)^3},\nonumber\\
u_3=\frac{\mu_3(2\mu_1+2\mu_2+\mu_3+\mu_4)(\mu_1+2\mu_2+2\mu_3+\mu_4)^2}{(\mu_2+\mu_3)(\mu_1+\mu_2+\mu_3+\mu_4)^3},\nonumber\\
u_4=\frac{\mu_4(2\mu_1+2\mu_2+\mu_3+\mu_4)(\mu_1+2\mu_2+2\mu_3+\mu_4)}{(\mu_1+\mu_2+\mu_3+\mu_4)^3},\label{Equis}
\end{gather}
with
\begin{displaymath}
 R_1=4,\qquad R_2=16,\qquad R_3=4,\qquad R_4=1.
 \end{displaymath}
The convergence region of the four-variable hypergeometric function $L$ needs to be projected to two variables. The function $H_d^{(p,q;r;s,t)}(u,v)$ in~\eqref{EqH} converges for those parameters that satisfy $(x_1=|u|,x_2=|u|,x_3=|u|,x_4=|v|)\in D$. The expressions for $u_i$'s in~\eqref{Equis} are too complicated to obtain an explicit form of the convergence region of $H_d^{(p,q;r;s,t)}(u,v)$, but convergence for any specific values of its parameters can be verified straightforwardly. The region of convergence is clearly finite, but not zero.

\subsection{Recurrence relation and symmetry}

In \cite{Fortin:2016dlj}, the recurrence relation
\begin{gather}
P_d^{(p+1,q;r;s,t)}(m,n) =\frac{r-s+p+1-d/2+m}{r-s+p+1-d/2}P_d^{(p,q;r;s,t)}(m,n)\nonumber\\
\qquad{}-\frac{(r-s-1+2m+n)(r-s-d/2+m)(r-t-1+2m+n)(r-t-d/2+m)}{r-s+p+1-d/2}\nonumber\\
\qquad{}\times P_d^{(p,q;r;s,t)}(m-1,n+1)\nonumber\\
\qquad{}+\frac{(r+m+n)(r-s-d/2+m)(r-t-d/2+m)(r-t+q+m+n)}{r-s+p+1-d/2}\nonumber\\
\qquad{}\times P_d^{(p,q;r;s,t)}(m-1,n+2),\label{EqRec}
\end{gather}
necessary to show that \eqref{EqH} is the appropriate solution to \eqref{EqGH} and the symmetry $H_d^{(p,q;r;s,t)}(u,v)\allowbreak =H_d^{(q,p;r;t,s)}(u,v)$ needed for the invariance under the dihedral group $D_6$ of order $12$ were not explicitly demonstrated to follow from the solution~\eqref{EqH}.

In the next sections, several equivalent expressions for $P_d^{(p,q;r;s,t)}(m,n)$ and $H_d^{(p,q;r;s,t)}(u,v)$ will be introduced to verify that~\eqref{EqRec} is satisfied and the solution~\eqref{EqH} is indeed invariant under~$D_6$.

\section[Several expressions for H]{Several expressions for $\boldsymbol{H}$}\label{SecExp}

In this section several equivalent but completely different expressions for the $H$-function are given. The first subsection lists the various expressions, while the proofs are left for the following subsections. The reader only interested in the different forms of $H$ can certainly skip the proofs.

\subsection[H-function]{$\boldsymbol{H}$-function}

By trivially combining Pochhammer symbols together, the original solution \eqref{EqH} for the $H$-function can be rewritten as
\begin{gather}
P_d^{(p,q;r;s,t)}(m,n) =\sum_{i,j\geq0}\frac{(-1)^i(-m)_i(-i)_j}{i!j!m!}(-p,r-t+1-d/2+m-i)_i\nonumber\\
\qquad{} \times (-s-q+m-i+j,r-t+2m+n-i+j)_{i-j}\nonumber\\
\qquad{} \times(r-s+m+n+i,r-s+p+1-d/2)_{m-i} (-q,-t)_{m-i+j}\nonumber\\
 \quad{} =\frac{(-q,-t,r-s+p+1-d/2,r-s+m+n)_m}{m!}\nonumber\\
\qquad{} \times F_{2,2,0}^{3,2,0}\left[\left.\begin{matrix}-m,-p,-r+t+d/2-m;s+q-m+1,-r+t-2m-n+1;-\\-r+s-p+d/2-m,r-s+m+n;q-m+1,t-m+1;- \end{matrix}\right|-1,1\right], \nonumber\\
H_d^{(p,q;r;s,t)}(u,v) =\sum_{i,j,m,n\geq0}\frac{(-1)^i(-m)_i(-i)_j}{i!j!m!n!}(-p)_i(-q,-t)_{m-i+j}(-s-q+m-i+j)_{i-j}\nonumber\\
\qquad{} \times (r-s+p+1-d/2)_{m-i}\nonumber\\
\qquad{} \times\frac{(r,r-s+p,r-t+q)_{m+n}}{(r-s)_{m+n+i}(r-s+1-d/2)_m(r-t)_{2m+n-i+j}(r-t+1-d/2)_{m-i}}x^my^n.\label{EqH1}
\end{gather}
As indicated above, $P_d^{(p,q;r;s,t)}(m,n)$ can also be expressed in terms of a Kamp\'e de F\'eriet function. This form leads to an alternative way of proving equivalences of different forms for the $H$-function.

Another expression for the $H$-function, which allows to show that $P_d^{(p,q;r;s,t)}(m,n)$ is invariant under the interchange of $r+1-d/2$ and $r+m+n$, is given by
\begin{gather}
P_d^{(p,q;r;s,t)}(m,n) =\sum_{i,j\geq0}\frac{(-m)_i(-i)_j}{i!j!m!}(-p)_{m-j}(-q,-t)_i\nonumber\\
\qquad\times{} (r-s+m-j+1-d/2,r-s+2m+n-j)_j(2r-s-t+2m+n-d/2)_{i-j} \nonumber\\
\qquad\times{} (-s-q+i,r-t+i+1-d/2,r-t+m+n+i)_{m-i} \nonumber\\
 \quad{} =\frac{(-p,-s-q,r-t+1-d/2,r-t+m+n)_m}{m!}\nonumber\\
 \qquad{} \times F_{3,0,1}^{3,1,2}\left[\left.\begin{matrix}-m,-q,-t;2r-s-t+2m+n-d/2;\\ -r+s-m+d/2,-r+s-2m-n+1\\-s-q,r-t+1-d/2,r-t+m+n;-;p-m+1\end{matrix}\right|1,1\right], \nonumber\\
H_d^{(p,q;r;s,t)}(u,v) =\sum_{i,j,m,n\geq0}\frac{(-m)_i(-i)_j}{i!j!m!n!}(-p)_{m-j}(-q,-t)_i(-s-q+i)_{m-i}\nonumber\\
\qquad{}\times (2r-s-t+2m+n-d/2)_{i-j}\nonumber\\
\qquad{} \times\frac{(r,r-s+p,r-t+q)_{m+n}}{(r-s)_{2m+n-j}(r-s+1-d/2)_{m-j}(r-t)_{m+n+i}(r-t+1-d/2)_i}x^my^n.\label{EqH2}
\end{gather}

A slightly more complicated expression with one extra sum, useful to prove the symmetry property $H_d^{(p,q;r;s,t)}(u,v)=H_d^{(q,p;r;t,s)}(u,v)$, corresponds to
\begin{gather}
P_d^{(p,q;r;s,t)}(m,n) =\sum_{i,j,k\geq0}\frac{(-m)_i(-i)_j(-i+j)_k}{i!j!k!m!}(-p)_{m-j}(-q)_i(-s)_{m-j-k}(-t)_{i-k}\nonumber\\
\qquad{}\times (r+1-d/2,r+m+n)_k(r-s+m-j+1-d/2,r-s+2m+n-j)_j\nonumber\\
\qquad{} \times(r-t+i+1-d/2,r-t+m+n+i)_{m-i},\nonumber\\
H_d^{(p,q;r;s,t)}(u,v) =\sum_{i,j,k,m,n\geq0}\frac{(-m)_i(-i)_j(-i+j)_k}{i!j!k!m!n!}(-p)_{m-j}(-q)_i(-s)_{m-j-k}\nonumber\\
\qquad{}\times (-t)_{i-k}(r+1-d/2)_k\nonumber\\
\qquad{} \times\frac{(r)_{m+n+k}(r-s+p,r-t+q)_{m+n}}{(r-s)_{2m+n-j}(r-s+1-d/2)_{m-j}(r-t)_{m+n+i}(r-t+1-d/2)_i}x^my^n.\label{EqH3}
\end{gather}
In this case, $P_d^{(p,q;r;s,t)}(m,n)$ is not expressible in terms of a Kamp\'e de F\'eriet function, but can be written as a generalized Lauricella function. As this way of expressing $P_d^{(p,q;r;s,t)}(m,n)$ does not lead to an alternative proof we do not provide it here.

The final rewriting of the $H$-function, relevant to prove the recurrence relation \eqref{EqRec}, is
\begin{gather}
P_d^{(p,q;r;s,t)}(m,n) =\sum_{i,j\geq0}\frac{(-m)_i(-i)_j}{i!j!m!}(-p,r-t+j+1-d/2)_{m-j}\nonumber\\
\qquad{} \times (-q,-t,r-s+p+1-d/2)_j
(r+m+n,r-t+q+m+n)_{i-j}\nonumber\\
\qquad{} \times (r-s+m+n+i-j)_{m-i+j}(r-t+m+n+i)_{m-i}\nonumber\\
\quad{} =\frac{(-p,r-s+m+n,r-t+1-d/2,r-t+m+n)_m}{m!}\nonumber\\
\qquad{} \times F_{1,1,2}^{1,2,3}\left[\left.\begin{matrix}-m;r+m+n,r-t+q+m+n;-q,-t,r-s+p+1-d/2\\r-t+m+n;r-s+m+n;p-m+1,r-t+1-d/2 \end{matrix}\right|1,1\right], \nonumber\\
H_d^{(p,q;r;s,t)}(u,v) =\sum_{i,j,m,n\geq0}\frac{(-m)_i(-i)_j}{i!j!m!n!}(-p)_{m-j}(-q,-t,r-s+p+1-d/2)_j
\nonumber\\
\qquad{} \times\frac{(r-s+p)_{m+n}(r,r-t+q)_{m+n+i-j}}{(r-s)_{m+n+i-j}(r-s+1-d/2)_m(r-t)_{m+n+i}(r-t+1-d/2)_j}x^my^n.\label{EqH4}
\end{gather}

\subsection{Proof of (\ref{EqH2})}

To prove \eqref{EqH2} from \eqref{EqH1}, it is convenient to reorder the sums in the polynomial using
\begin{displaymath}
\sum_{i=0}^m\sum_{j=0}^ia_{ij}=\sum_{i=0}^m\sum_{j=0}^ia_{m-j,i-j} ,
\end{displaymath}
which leads to
\begin{gather}
P_d^{(p,q;r;s,t)}(m,n) =\sum_{i=0}^m\sum_{j=0}^i\frac{(-m)_i(-i)_j}{i!j!m!} (-q,-t)_i \nonumber\\
\qquad{}\times (r-s+2m+n-j,r-s+p+1-d/2)_j(-p,r-t+j+1-d/2)_{m-j}\nonumber\\
\qquad{} \times (-s-q+i,r-t+m+n+i)_{m-i},\label{EqP}
\end{gather}
after simplifying the pre-factors. With the help of
\begin{gather*}
(-p)_{m-j} =\frac{(-1)^j(-p)_m}{(p-m+1)_j},\\
(r-s+2m+n-j)_j =(-1)^j(s-r-2m-n+1)_j,\\
(r-t+1-d/2+j)_{m-j} =\frac{(r-t+1-d/2)_m}{(r-t+1-d/2)_j}
\end{gather*}
and the first symmetry of \eqref{EqSym} on the sum over $j$ with $a=s-r-2m-n+1$ and $d=r-t+1-d/2$, one finally obtains \eqref{EqH2} after some trivial simplifications. Hence, $P_d^{(p,q;r;s,t)}(m,n)$ is invariant under the interchange of $r+1-d/2$ and $r+m+n$.

An alternative to this derivation is to use the Kamp\'e de F\'eriet forms of $P_d^{(p,q;r;s,t)}(m,n)$ in~\eqref{EqH1} and \eqref{EqH2}.\footnote{This form of the proof was suggested by an anonymous referee.} The $F_{2,2,0}^{3,2,0}$ Kamp\'e de F\'eriet function in~\eqref{EqH1} can be re-expressed using the reversal of summation order for the hypergeometric series
\begin{gather*}
F_{a,b,0}^{a+1,b,0}\left[\left.\begin{matrix}-m,\alpha_1,\ldots,\alpha_a;\beta_1,\ldots,\beta_b;-\\ \gamma_1,\ldots,\gamma_a;\delta_1,\ldots,\delta_b;-\end{matrix}\right|x,y\right]
=(-x)^m\frac{(\alpha_1,\ldots,\alpha_a;\beta_a,\ldots,\beta_b)_m}{(\gamma_1,\ldots,\gamma_a;\delta_1,\ldots,\delta_b)_m}\qquad\qquad\\
 \qquad{}\times F_{b,0,a}^{b+1,0,a}\!\left[\left.\begin{matrix}-m,1-m-\delta_1,\ldots,1-m-\delta_b;-,1-m-\gamma_1, \ldots,1-m-\gamma_a\\1-m-\beta_1,\ldots,1-m-\beta_b;-;1-m-\alpha_1,\ldots,1-m-\alpha_a\end{matrix}\right|-\frac{y}{x},\frac{1}{x}\right],
\end{gather*}
yielding
\begin{gather}
 F_{2,2,0}^{3,2,0}\left[\left.\begin{matrix}-m,-p,-r+t+d/2-m;s+q-m+1,-r+t-2m-n+1;-\\-r+s-p+d/2-m,r-s+m+n;q-m+1,t-m+1;-\end{matrix} \right|-1,1\right]\nonumber\\
 \quad{}=\frac{(-p,-r+t+d/2-m,s+q-m+1,-r+t-2m-n+1)_m}{(-r+s-p+d/2-m,r-s+m+n,q-m+1,t-m+1)_m }\nonumber\\
 \qquad{}\times F_{2,0,2}^{3,0,2} \left[\left.\begin{matrix}-m,-q,-t;-;1+r-s+p-d/2,1-r+s-2m-n\\-s-q,r-t+m+n;-;1-m+p,1+r-t-d/2 \end{matrix}\right| 1,1\right].\label{EqKdFreverse}
\end{gather}

Theorem~3.5 in \cite{Srivastava2013} states that for any sequence of complex numbers $\Omega_k$
\begin{gather*}
 \sum_{i,j\geq0}\frac{z^i}{i!}\frac{(\alpha)_j(\beta)_j(-z)^j}{(\gamma)_j(\delta)_jj!} (i+j)!\Omega_{i+j}=\sum_{i,j\geq0}\frac{(\delta-\beta)_iz^i}{i!}\frac{(\gamma-\alpha)_j(\beta)_j z^j}{j!} \frac{(i+j)!}{(\gamma)_j(\delta)_{i+j}}\Omega_{i+j}.
\end{gather*}
Setting $z=1$ in the above result and substituting
\begin{gather*}
\alpha\to1+r-s-p-d/2,\qquad\beta\to1-r+s-2m-n,\qquad\gamma\to1-m +p,\\
\delta\to1+r-t-d/2,\qquad\Omega_k\to\frac{(-m,-q,-t)_k}{k!(-s-q,r-t+m+n)_k},
\end{gather*}
transforms the $F_{2,0,2}^{3,0,2}$ Kamp\'e de F\'eriet on the right-hand side of \eqref{EqKdFreverse} to $F_{3,0,1}^{3,1,2}$ in~\eqref{EqH2}.

\subsection{Proof of (\ref{EqH3})}

Now that the equivalence of \eqref{EqH2} and \eqref{EqH1} is established, the third form for $P_d^{(p,q;r;s,t)}(m,n)$ can be obtained from \eqref{EqH2}. Using the binomial identity \eqref{EqBinom} in \eqref{EqH2} to express
\begin{displaymath}
(-s-q+i)_{m-i}=\sum_{k=0}^{m-i}\genfrac{(}{)}{0pt}{0}{m-i}{k}(-s)_{m-i-k}(-q+i)_k =\sum_{k=i}^m\genfrac{(}{)}{0pt}{0}{m-i}{k-i}(-s)_{m-k}(-q+i)_{k-i}
\end{displaymath}
allows to combine the last Pochhammer symbol above with $(-q)_i$ in \eqref{EqH2}, leading to
\begin{gather*}
P_d^{(p,q;r;s,t)}(m,n) =\sum_{i=0}^m\sum_{j=0}^i\sum_{k=i}^m\genfrac{(}{)}{0pt}{0}{m-i}{k-i}\frac{(-m)_i(-i)_j}{i!j!m!}(-p)_{m-j}(-q)_k(-s)_{m-k}(-t)_i \\
\qquad{}\times (2r-s-t+2m+n-d/2)_{i-j}(r-s+m-j+1-d/2,r-s+2m+n-j)_j\\
\qquad{} \times(r-t+i+1-d/2,r-t+m+n+i)_{m-i}.
\end{gather*}
Reordering the sums as
\begin{displaymath}
 \sum_{i=0}^m\sum_{j=0}^i\sum_{k=i}^ma_{ijk}=\sum_{k=0}^m\sum_{j=0}^k\sum_{i=0}^{k-j}a_{i+j,jk}\,,
\end{displaymath}
the previous result becomes
\begin{gather*}
P_d^{(p,q;r;s,t)}(m,n) =\sum_{k=0}^m\sum_{j=0}^k\sum_{i=0}^{k-j}\genfrac{(}{)}{0pt}{0}{m}{k}\genfrac{(}{)}{0pt}{0}{k}{j}\frac{(-k+j)_i}{i!m!} (-p)_{m-j}(-q)_k(-s)_{m-k}(-t)_{i+j}\\
\qquad{} \times(2r-s-t+2m+n-d/2)_i(r-s+m-j+1-d/2,r-s+2m+n-j)_j \\
\qquad{} \times(r-t+i+j+1-d/2,r-t+m+n+i+j)_{m-i-j},
\end{gather*}
after a trivial simplification of the pre-factors. Using
\begin{gather*}
(-t)_{i+j} =(-t)_j(-t+j)_i,\\
(r-t+i+j+1-d/2)_{m-i-j} =\frac{(r-t+j+1-d/2)_{m-j}}{(r-t+j+1-d/2)_i},\\
(r-t+m+n+i+j)_{m-i-j} =\frac{(r-t+m+n+j)_{m-j}}{(r-t+m+n+j)_i},
\end{gather*}
and separating the sum over $i$ gives
\begin{gather*}
P_d^{(p,q;r;s,t)}(m,n) =\sum_{k=0}^m\sum_{j=0}^k\frac{1}{m!}\genfrac{(}{)}{0pt}{0}{m}{k}\genfrac{(}{)}{0pt}{0}{k}{j}(-q)_k\\
\qquad{}\times (-t,r-s+m-j+1-d/2,r-s+2m+n-j)_j\\
\qquad{} \times (-p)_{m-j} (-s)_{m-k} (r-t+j+1-d/2,r-t+m+n+j)_{m-j}\\
\qquad{} \times\sum_{i=0}^{k-j}\frac{(-k+j)_i}{i!}\frac{(-t+j,2r-s-t+2m+n-d/2)_i}{(r-t+j+1-d/2,r-t+m+n+j)_i}.
\end{gather*}
At this point, the last symmetry property \eqref{EqSym} with $a=-t+j$ can be used for the sum over $i$ leading to
\begin{gather*}
P_d^{(p,q;r;s,t)}(m,n) =\sum_{k=0}^m\sum_{j=0}^k\frac{1}{m!}\genfrac{(}{)}{0pt}{0}{m}{k}\genfrac{(}{)}{0pt}{0}{k}{j}(-q)_k\\
\qquad{}\times (-t,r-s+m-j+1-d/2,r-s+2m+n-j)_j\\
\qquad{} \times (-p)_{m-j} (-s)_{m-k} (r-t+j+1-d/2,r-t+m+n+j)_{m-j}\\
\qquad{} \times\frac{(-t+j,s-m+j+1)_{k-j}}{(r-t+j+1-d/2,r-t+m+n+j)_{k-j}} \\
\qquad{}\times \sum_{i=0}^{k-j}\frac{(-k+j)_i}{i!}\frac{(r+1-d/2,r+m+n)_i}{(s-m+j+1,t-k+1)_i}.
\end{gather*}
Combining the Pochhammer symbols in the last line yields
\begin{gather*}
P_d^{(p,q;r;s,t)}(m,n) =\sum_{k=0}^m\sum_{j=0}^k\frac{1}{m!}\genfrac{(}{)}{0pt}{0}{m}{k}\genfrac{(}{)}{0pt}{0}{k}{j}(-q)_k\\
\qquad{}\times (-t,r-s+m-j+1-d/2,r-s+2m+n-j)_j\\
\qquad{} \times (-p)_{m-j} (-s)_{m-k} (r-t+j+1-d/2,r-t+m+n+j)_{m-j}\\
\qquad{} \times (-1)^{k+j}\sum_{i=0}^{k-j}\frac{(-k+j)_i}{i!}\frac{(r+1-d/2,r+m+n)_i(-t+j,-s+m-k)_{k-j-i}}{(r-t+j+1-d/2,r-t+m+n+j)_{k-j}},
\end{gather*}
which is equivalent to \eqref{EqH3} once a few simplifications of the Pochhammer symbols are performed and the indices are changed as in $i\leftrightarrow k$.

\subsection{Proof of (\ref{EqH4})}

The expression \eqref{EqH4} can be obtained starting from the form \eqref{EqP} which can be written as
\begin{gather*}
P_d^{(p,q;r;s,t)}(m,n) =\sum_{i=0}^m\sum_{j=0}^i\frac{(-1)^{i+j}}{(m-i)!(i-j)!j!} (-q,-t)_i\\
\qquad{}\times (r-s+2m+n-j,r-s+p+1-d/2)_j\\
\qquad{} \times(-s-q+i,r-t+m+n+i)_{m-i} (-p,r-t+j+1-d/2)_{m-j},
\end{gather*}
after simplifying the pre-factors. Reordering the sums using
\begin{displaymath}
 \sum_{i=0}^m\sum_{j=0}^ia_{ij}=\sum_{j=0}^m\sum_{i=0}^{m-j}a_{i+j,j}\,,
\end{displaymath}
leads to
\begin{gather*}
P_d^{(p,q;r;s,t)}(m,n) =\sum_{j=0}^m\frac{1}{(m-j)!j!} (-q,-t,r-s+2m+n-j,r-s+p+1-d/2)_j\\
\qquad{} \times (-p, -s-q+j, r-t+m+n+j,r-t+j+1-d/2)_{m-j}\\
\qquad{} \times\sum_{i=0}^{m-j}\frac{(-m+j)_i}{i!}\frac{(-q+j,-t+j)_i}{(-s-q+j,r-t+m+n+j)_i},
\end{gather*}
where the sum over $i$ was factored out and its pre-factor simplified. Using the second symmetry property \eqref{EqSym} for the sum over $i$ with $d=-s-q+j$ gives
\begin{gather*}
P_d^{(p,q;r;s,t)}(m,n) =\sum_{j=0}^m\frac{1}{(m-j)!j!} (-q, -t, r-s+2m+n-j,r-s+p+1-d/2)_j\\
\qquad{} \times (-p,-s-q+j,r-t+m+n+j,r-t+j+1-d/2)_{m-j}\\
\qquad{} \times\frac{(r-s+m+n)_{m-j}}{(-s-q+j)_{m-j}}\sum_{i=0}^{m-j}\frac{(-m+j)_i}{i!}\frac{(r-t+q+m+n,r+m+n)_i}{(r-t+m+n+j,r-s+m+n)_i}.
\end{gather*}
Combining the Pochhammer symbols together leads to
\begin{gather*}
P_d^{(p,q;r;s,t)}(m,n) =\sum_{j=0}^m\sum_{i=0}^{m-j}\frac{(-m)_{i+j}(-i-j)_j}{(i+j)!j!m!}(-p)_{m-j}(-q,-t)_j(r-s+m+n+i)_{m-i}\\
\qquad{} \times(r-s+p+1-d/2)_j(r-t+m+n+i+j)_{m-j-i}\\
\qquad{} \times (r-t+j+1-d/2)_{m-j} (r-t+q+m+n,r+m+n)_i,
\end{gather*}
after straightforward simplifications of the pre-factors. Shifting $i\to i-j$ followed by reversing the order of the sums brings the results to
\begin{gather*}
P_d^{(p,q;r;s,t)}(m,n) =\sum_{i=0}^m\sum_{j=0}^i\frac{(-m)_i(-i)_j}{i!j!m!}(-p)_{m-j}(-q,-t)_j(r-s+m+n+i-j)_{m-i+j}\\
\qquad{} \times(r-s+p+1-d/2)_j(r-t+m+n+i)_{m-i}(r-t+j+1-d/2)_{m-j}\\
\qquad{} \times(r-t+q+m+n,r+m+n)_{i-j},
\end{gather*}
which is exactly \eqref{EqH4}.

\section{Recurrence relation and symmetry}\label{SecRec}

This section proves the recurrence relation and the symmetry using suitable expressions for the $H$-function obtained in the previous section.

\subsection{Proof of the symmetry}

In \cite{Fortin:2016dlj} it was argued from the definition of the $H$-function in terms of the differential opera\-tor~\eqref{EqGH} that $H_d^{(p,q;r;s,t)}(u,v)=H_d^{(q,p;r;t,s)}(u,v)$, a symmetry property necessary to show that the $H$-function is invariant under $D_6$. At the level of the polynomial $P_d^{(p,q;r;s,t)}(m,n)$, the previous symmetry corresponds simply to $P_d^{(p,q;r;s,t)}(m,n)=P_d^{(q,p;r;t,s)}(m,n)$. It is trivial to show this property directly using expression \eqref{EqH3}.

Indeed, \eqref{EqH3} implies that
\begin{gather*}
P_d^{(q,p;r;t,s)}(m,n) =\sum_{i=0}^m\sum_{j=0}^i\sum_{k=0}^{i-j}\frac{(-m)_i(-i)_j(-i+j)_k}{i!j!k!m!}(-q)_{m-j}(-p)_i(-t)_{m-j-k}(-s)_{i-k}\\
\qquad{}\times (r+1-d/2,r+m+n)_k(r-t+m-j+1-d/2,r-t+2m+n-j)_j\\
\qquad{} \times(r-s+i+1-d/2,r-s+m+n+i)_{m-i}.
\end{gather*}
Since
\begin{displaymath}
 \sum_{i=0}^m\sum_{j=0}^ia_{ij}=\sum_{i=0}^m\sum_{j=0}^ia_{m-j,m-i},
\end{displaymath}
the previous result can be rewritten as
\begin{gather*}
P_d^{(q,p;r;t,s)}(m,n) =\sum_{i=0}^m\sum_{j=0}^i\sum_{k=0}^{i-j}\frac{(-m)_i(-i)_j(-i+j)_k}{i!j!k!m!}(-q)_i(-p)_{m-j} (-t)_{i-k}(-s)_{m-j-k}\\
\qquad{}\times (r+1-d/2,r+m+n)_k (r-t+i+1-d/2,r-t+m+n+i)_{m-i}\\
\qquad{}\times(r-s+m-j+1-d/2,r-s+2m+n-j)_j
 =P_d^{(p,q;r;s,t)}(m,n),
\end{gather*}
where the pre-factors in the first equality have been simplified. This result for $P_d^{(p,q;r;s,t)}(m,n)$ therefore shows that the $H$-function is invariant under the dihedral group of order~$12$.

\subsection{Proof of the recurrence relation}

The recurrence relation \eqref{EqRec} can be verified directly starting from expression \eqref{EqH4}. It is actually simpler to introduce a generalization of \eqref{EqH4} in order to prove \eqref{EqRec}. Defining
\begin{equation} \label{EqQ}
 Q_d^{(p,q,t,a,b,c,d,e,f)}(m)=\sum_{i=0}^m\sum_{j=0}^i\frac{(-m)_i(-i)_j}{i!j!m!}\frac{(-p)_{m-j}(-q,-t,a)_j(b,c)_{i-j}}{(d)_j(e)_{i-j}(f)_i},
\end{equation}
the original polynomial can be written as
\begin{gather}
P_d^{(p,q;r;s,t)}(m,n)\label{EqPQ}\\
\qquad{} =(d,e,f)_m Q_d^{(p,q,t,r-s+p+1-d/2,r+m+n,r-t+q+m+n,r-t+1-d/2,r-s+m+n,r-t+m+n)}(m).\nonumber
\end{gather}
The new polynomial $Q$ satisfies several contiguous relations. Two such relations are needed to prove \eqref{EqRec}. Using the fact that
\begin{gather*}
(-p)_{m-j} =(-p-1)_{m-j}+(m-i)(-p)_{m-1-j}+(i-j)(-p)_{m-1-j},\\
a(a+1)_j =(a+m)(a)_j-(m-i)(a)_j-(i-j)(a)_j,
\end{gather*}
leads directly to the two following contiguous relations for \eqref{EqQ},
\begin{gather*}
Q_d^{(p,q,t,a,b,c,d,e,f)}(m) =Q_d^{(p+1,q,t,a,b,c,d,e,f)}(m)+Q_d^{(p,q,t,a,b,c,d,e,f)}(m-1)\\
\hphantom{Q_d^{(p,q,t,a,b,c,d,e,f)}(m) =}{} -\frac{bc}{ef}Q_d^{(p,q,t,a,b+1,c+1,d,e+1,f+1)}(m-1),\\
aQ_d^{(p+1,q,t,a+1,b,c,d,e,f)}(m) =(a+m)Q_d^{(p+1,q,t,a,b,c,d,e,f)}(m)+(p+1)Q_d^{(p,q,t,a,b,c,d,e,f)}(m-1)\\
\hphantom{aQ_d^{(p+1,q,t,a+1,b,c,d,e,f)}(m) =}{} -\frac{bc}{ef}(p+1)Q_d^{(p,q,t,a,b+1,c+1,d,e+1,f+1)}(m-1).
\end{gather*}
These two contiguous relations are not obeyed by the polynomial $P_d^{(p,q;r;s,t)}(m,n)$ due to the relationship between the different parameters in \eqref{EqPQ}. However, by isolating $Q_d^{(p+1,q,t,a,b,c,d,e,f)}(m)$ in the first contiguous relation and inserting its definition in the right-hand side of the second contiguous relation, it is easy to obtain the following contiguous relation
\begin{gather}
aQ_d^{(p+1,q,t,a+1,b,c,d,e,f)}(m) \nonumber\\
\quad{}=(a+m)Q_d^{(p,q,t,a,b,c,d,e,f)}(m)-(a-p+m-1)Q_d^{(p,q,t,a,b,c,d,e,f)}(m-1)\nonumber\\
\quad\quad{} +\frac{bc}{ef}(a-p+m-1)Q_d^{(p,q,t,a,b+1,c+1,d,e+1,f+1)}(m-1),\label{EqCR}
\end{gather}
which is satisfied by the polynomial $P_d^{(p,q;r;s,t)}(m,n)$. In fact, \eqref{EqCR} is nothing else than the recurrence relation \eqref{EqRec}, proving that the $H$-function is the correct quantity appearing in conformal blocks.

\section{Differential operators}\label{SecDiff}

In this section the differential operator $\mathcal{D}_{(u,v)}$ is used to derive both the $G$ and $H$-functions constructively. Generalizations to higher-point correlation functions will be discussed elsewhere.

\subsection{Action}

By direct computation, the action of the differential operator $\mathcal{D}_{(u,v)}$ on the variables $x=u/v$ and $y=1-1/v$ is simply
\begin{displaymath}
 \mathcal{D}_{(u,v)}x^my^n=(-2)\big[n(n-1)x-n(m+n)xy+(m+n)(m+1-d/2)y^2\big]x^{m+1}y^{n-2},
\end{displaymath}
and therefore
\begin{gather}
\mathcal{D}_{(u,v)}^hx^my^n =(-2)^h\sum_{i,j\geq0}\frac{(-1)^{i+j}(-h)_i(-i)_j}{i!j!}(-n)_{i+j}(m+i+1-d/2)_{h-i}\nonumber\\
\hphantom{\mathcal{D}_{(u,v)}^hx^my^n =}{} \times (m+n)_{h-j}x^{m+h+i}y^{n-i-j}.\label{EqAction}
\end{gather}
Expression \eqref{EqAction} has the correct limiting behavior at $(u,v)\to(0,1)$ as can be checked by computing \eqref{EqG} from \eqref{EqGH}, which gives
\begin{align*}
G_d^{(q;r;t)}(u,v)&=\frac{x^{-(r-t+q)}\mathcal{D}_{(u,v)}^qx^{r-t}(1-y)^{-r}}{(-2)^q(r-t,r-t+1-d/2)_q}=\sum_{k\geq0}\frac{(r)_k}{k!}\frac{x^{-(r-t+q)}\mathcal{D}_{(u,v)}^qx^{r-t}y^k}{(-2)^q(r-t,r-t+1-d/2)_q}\\
&=\sum_{i,j,k\geq0}\frac{(-1)^{i+j}(-i)_j(-k)_{i+j}}{i!j!k!}\frac{(-q)_i(r)_k(r-t+k)_{q-j}}{(r-t)_q(r-t+1-d/2)_i}x^iy^{k-i-j}\\
&=\sum_{m,n,j\geq0}\frac{(-m)_j}{j!}\frac{(r+m+n)_j}{(r-t+m+n)_j}\frac{(-q)_m(r)_{n+m}(r-t+m+n)_q}{(r-t)_q(r-t+1-d/2)_mm!n!}x^my^n\\
&=\sum_{m,n\geq0}\frac{(-t)_m}{(r-t+m+n)_m}\frac{(-q)_m(r)_{n+m}(r-t+m+n)_q}{(r-t)_q(r-t+1-d/2)_mm!n!}x^my^n.
\end{align*}
In the third equality the sums where shifted $(i,k)\to(m,n+i+j)$, while the Vandermonde's identity \eqref{EqV} was used in the last equality. The final result is equivalent to \eqref{EqG} and thus proves that \eqref{EqAction} is the correct action of the differential operator in computing conformal blocks.

Using \eqref{EqAction}, the $H$-function can be easily computed from \eqref{EqGH}
\begin{gather*}
H_d^{(p,q;r;s,t)}(u,v) =\frac{\left(\frac{u}{v}\right)^{-(r-s+p)}\mathcal{D}_{(u,v)}^p\left(\frac{u}{v}\right)^{r-s}G_d^{(q;r;t)}(u,v)}{(-2)^p(r-s,r-s+1-d/2)_p}\\
 =\sum_{m,n\geq0}\frac{(-q,-t)_m}{(r-t+1-d/2)_mm!}\frac{(r,r-t+q)_{m+n}}{(r-t)_{2m+n}n!}\frac{x^{-(r-s+p)}\mathcal{D}_{(u,v)}^px^{r-s+m}y^n}{(-2)^p(r-s,r-s+1-d/2)_p}\\
 =\sum_{i,j,m,n\geq0}\frac{(-1)^{i+j}(-p)_i(-i)_j}{i!j!}\frac{(-q,-t)_m}{(r-t+1-d/2)_mm!}\frac{(r,r-t+q)_{m+n}}{(r-t)_{2m+n}n!}\\
 \quad{}\times\frac{(-n)_{i+j}(r-s+m+i+1-d/2)_{p-i}(r-s+m+n)_{p-j}}{(r-s,r-s+1-d/2)_p}x^{m+i}y^{n-i-j}\\
 =\sum_{i,j,m,n\geq0}\frac{(-1)^{i+j}(-p)_i(-i)_j}{i!j!}\frac{(-q,-t)_{m-i}}{(r-t+1-d/2)_{m-i}(m-i)!}\frac{(r,r-t+q)_{m+n+j}}{(r-t)_{2m+n-i+j}(n+i+j)!}\\
\quad{} \times\frac{(-n-i-j)_{i+j}(r-s+p+1-d/2)_{m-i}(r-s+p)_{m+n}}{(r-s)_{m+n+j}(r-s+1-d/2)_m}x^my^n,
\end{gather*}
where the last identity, obtained by substituting $(m,n)\to(m-i,n+i+j)$, corresponds exactly to \eqref{EqH4} after changing $(i,j)\to(m-j,j-i)$. Since \eqref{EqH4} is the expression for the $H$-function that originates directly from the action of $\mathcal{D}_{(u,v)}$, it is now clear why~\eqref{EqH4} is the appropriate form to prove the recurrence relation~\eqref{EqRec}.

\section{Conclusion}\label{SecConc}

We used several identities for the Pochhammer symbols and hypergeometric-like polynomials in order to show that the $H$-function computed in \cite{Fortin:2016dlj} is the appropriate function appearing in conformal blocks. With the help of these identities, several different expressions for the $H$-function were presented. This allowed us to demonstrate explicitly that the $H$-function is invariant under the dihedral group of order $12$ and that it satisfies the proper recurrence relation.

We also found the explicit action of the differential operator on simple products of the conformal cross-ratios. This differential form was used to give a constructive proof of the $H$-function, independent of the approach based on identities used before. As far as computing conformal blocks is concerned, the action of the differential operator is actually the most important result of this paper. Indeed, there exists a generalization of this expression that acts straightforwardly on higher $N$-point correlation functions. This result will be discussed elsewhere.

Finally, it is worth mentioning that the physical interpretation behind the $D_6$-symmetry of the $H$-function remains unclear. Nevertheless, this symmetry might have implications for the analyticity properties in spin of the conformal blocks.

\appendix

\section{Pochhammer symbols and hypergeometric functions}\label{SecPS}

Pochhammer symbols and hypergeometric functions satisfy several mathematical properties and some of those properties are necessary to show that the different representations of the $H$-function are equivalent. For completeness, this appendix presents several useful identities for the Pochhammer symbols and hypergeometric functions.

First, the Pochhammer symbol $(x)_\alpha$ is defined as
\begin{equation*} %\label{EqPS}
 (x)_\alpha=\frac{\Gamma(x+\alpha)}{\Gamma(x)},
\end{equation*}
and for $\alpha=n$ a non-negative integer, it satisfies
\begin{equation} \label{EqPSminus}
 (-x)_n=(-1)^n(x-n+1)_n,
\end{equation}
as well as the binomial identity
\begin{equation} \label{EqBinom}
 (x+y)_n=\sum_{k\geq0}^n \genfrac{(}{)}{0pt}{0}{n}{k}(x)_k(y)_{n-k}.
\end{equation}
The Vandermonde's identity
\begin{equation} \label{EqV}
(x+y)_\alpha=\sum_{k\geq0}\frac{(-\alpha)_k}{k!}(-x)_k(y+k)_{\alpha-k}.
\end{equation}
can be obtained from the binomial identity \eqref{EqBinom} by using \eqref{EqPSminus}. Unlike the binomial identity, the Vandermonde's identity is satisfied for any $\alpha$ (not just for integer values) as long as the sum converges. Both \eqref{EqBinom} and \eqref{EqV} can be demonstrated by recurrence.

Similarly, the ${}_3F_2$-hypergeometric functions of interest here satisfy
\begin{align}
{}_3F_2\left[\left.\begin{matrix}-n,a,b\\c,d\end{matrix}\right|1\right]&=\frac{(d-a)_n}{(d)_n}{}_3F_2 \left[\left.\begin{matrix}-n,a,c-b\\c,a-d+1-n\end{matrix}\right|1\right] \nonumber\\
&=\frac{(c+d-a-b)_n}{(d)_n}{}_3F_2\left[\left.\begin{matrix}-n,c-a,c-b\\c,c+d-a-b\end{matrix}\right|1\right]\nonumber\\
&=\frac{(a)_n(c+d-a-b)_n}{(c)_n(d)_n}{}_3F_2\left[\left.\begin{matrix}-n,d-a,c-a\\c+d-a-b,-a+1-n\end{matrix}\right|1\right].\label{EqSym}
\end{align}
These identities follow from the Thomae's relations for ${}_3F_2$-function with unit argument \cite{Gasper2004, Rainville1960}. Group theoretic origins of such identities were explored in~\cite{Beyer1987}, where it was shown that ${}_3F_2$-series with unit argument has an $S_5$ permutation invariance.

\subsection*{Acknowledgements}

Two of the authors (JFF and WS) would like to thank the CERN Theory Group, where this work was conceived, for its hospitality. The work of VC and JFF is supported by NSERC and FRQNT. We would like to acknowledge anonymous referees whose comments helped us improve the content and clarity of this article.

\pdfbookmark[1]{References}{ref}
\LastPageEnding

\end{document}